# Magnetism-induced massive Dirac spectra and topological defects in the surface state of Cr-doped $Bi_2Se_3$-bilayer topological insulators

C.-C. Chen[1,2], M. L. Teague[1,2], L. He[3], X. Kou[3], M. Lang[3], W. Fan[1], N. Woodward[1], K.-L. Wang[3] and N.-C. Yeh[1,2,4*]

[1] Department of Physics, California Institute of Technology, Pasadena, CA 91125, USA
[2] Institute for Quantum Information and Matter, California Institute of Technology, Pasadena, CA 91125, USA
[3] Department of Electrical Engineering, University of California, Los Angeles, CA 90095, USA
[4] Kavli Nanoscience Institute, California Institute of Technology, Pasadena, CA 91125, USA
[*] E-mail: ncyeh@caltech.edu

**Abstract.** Proximity-induced magnetic effects on the surface Dirac spectra of topological insulators are investigated by scanning tunneling spectroscopic (STS) studies of bilayer structures consisting of undoped $Bi_2Se_3$ thin films on top of Cr-doped $Bi_2Se_3$ layers. For thickness of the top $Bi_2Se_3$ layer equal to or smaller than 3 quintuple layers (QL), a spatially inhomogeneous surface spectral gap $\Delta$ opens up below a characteristic temperature $T_c^{2D}$, which is much higher than the bulk Curie temperature $T_c^{3D}$ determined from the anomalous Hall resistance. The mean value and spatial homogeneity of the gap $\Delta$ generally increase with increasing $c$-axis magnetic field ($H$) and increasing Cr doping level ($x$), suggesting that the physical origin of this surface gap is associated with proximity-induced $c$-axis ferromagnetism. On the other hand, the temperature ($T$) dependence of $\Delta$ is non-monotonic, showing initial increase below $T_c^{2D}$, which is followed by a "dip" and then rises again, reaching maximum at $T \ll T_c^{3D}$. These phenomena may be attributed to proximity magnetism induced by two types of contributions with different temperature dependences: a three-dimensional contribution from the bulk magnetism that dominates at low $T$, and a two-dimensional contribution associated with the RKKY interactions mediated by surface Dirac fermions, which dominates at $T_c^{3D} \ll T < T_c^{2D}$. In addition to the observed proximity magnetism, spatially localized sharp resonant spectra are found along the boundaries of gapped and gapless regions. These spectral resonances are long-lived at $H = 0$, with their occurrences being most prominent near $T_c^{2D}$ and becoming suppressed under strong $c$-axis magnetic fields. We attribute these phenomena to magnetic impurity-induced topological defects in the spin texture of surface Dirac fermions, with the magnetic impurities being isolated Cr impurities distributed near the interface of the bilayer system. The long-term stability of these topologically protected two-level states may find potential applications to quantum information technology.

## 1. Introduction

The research of topological matter is an exciting frontier where the classification of quantum states of matter beyond the principle of symmetry breaking has stimulated many conceptual advances and experimental discoveries [1-3]. Among various topological matter, topological insulators (TIs) [4-8] are bulk insulators in two or three dimensions with strong spin-orbit coupling and gapless surface states protected by the time-reversal invariance (TRI). The gapless surface state of TIs consists of an odd number of Dirac cones where the energy-momentum dispersion relation is linear, similar to the massless Dirac fermions in graphene except for the odd number of Dirac cones and an additional spin-momentum locking in the former. Gapping the Dirac cones of TIs by introducing superconductivity [9-11] or magnetism [8, 9, 12, 13] via either doping or proximity effects can provide feasible means to realize the elusive Majorana modes [9, 10, 14] and topological magnetoelectric (TME) effect [12, 13] in condensed matter systems.

The underlying physics for a magnetism-induced surface gap in three-dimensional (3D) TIs is based on the assumption that the long-range ferromagnetism has a net magnetization $M$ perpendicular to the surface of the 3D-TI, and that the in-plane component of the Hamiltonian $\mathcal{H}_0$ for the surface Dirac fermions remains intact after the introduction of c-axis magnetization. Hence, the total Hamiltonian $\mathcal{H}_{\text{total}}$ for the Dirac fermions becomes

$$\mathcal{H}_{\text{total}} = \mathcal{H}_0 + \frac{1}{2}J_{\text{eff}}M\sigma_z \equiv \hbar v_F\left(k_x\sigma_y - k_y\sigma_x\right) + \frac{1}{2}J_{\text{eff}}M\sigma_z \;, \qquad (1)$$

where $\sigma_{x,y,z}$ are the Pauli matrices, $k_{x,y}$ refer to the in-plane momentum of the Dirac fermions, $J_{\text{eff}}$ denotes the ferromagnetic coupling constant in the surface state, and $v_F$ is the Fermi velocity. The energy dispersion relation $E_k^{\pm}$ for the magnetism-induced massive Dirac fermions can be obtained by diagonalizing $\mathcal{H}_{\text{total}}$ in Eq. (1), which yields

$$E_k^{\pm} = \pm\sqrt{(v_F k)^2 + (J_{\text{ex}} M/2)^2} \;, \qquad (2)$$

where $k^2 \equiv k_x^2 + k_y^2$. Hence, an energy gap $\Delta = (J_{\text{eff}}M)$ opens up at the Dirac point for a finite c-axis magnetization $|M| > 0$ according to Eq. (2).

Experimental evidences for the occurrence of long-range ferromagnetism in the surface state of 3D-TIs have been manifested by the observation of surface gap opening in angle resolved photoemission spectroscopy (ARPES) [15-18] and the confirmation of quantized anomalous Hall effect (QAHE) [19,20] in magnetic 3D-TIs $(Bi_{1-x}Cr_x)_2Te_3$. However, the microscopic mechanism that mediates long-range ferromagnetism in the surface state has not been fully established [18, 21-26]. A number of puzzling phenomena, such as a surface gap opening at a temperature $T_c^{2D}$ much higher than the onset temperature of bulk magnetization $T_c^{3D}$ [15-17,22-25], the absence of gap formation by direct surface magnetic doping [27], and the lack of direct STS evidence for either magnetism-induced surface gaps [30] or spectroscopic magnetic impurity resonances [21], all suggest that further investigation is needed.

We report in this work direct evidence for magnetism-induced surface-state energy gaps and magnetic impurity resonances in 3D-TIs by STS studies of bilayer structures of $Bi_2Se_3$ and Cr-doped $Bi_2Se_3$. The bilayer samples were grown by molecular beam epitaxy (MBE) on InP (111) single crystalline substrates, with an updoped $Bi_2Se_3$ layer of varying thicknesses, $d_1$ quintuple layers (QLs), on top of a Cr-doped $Bi_2Se_3$ layer of a fixed thickness $d_2 = 6$-QL. These bilayer structures ensured that magnetism may be observed in the undoped $Bi_2Se_3$ through the proximity effect for sufficiently small $d_1$ values, which prevented possible complications due to Cr-doping induced changes in the electronic bandstructures of $Bi_2Se_3$, and thus enabled direct observation of the effect of global time-reversal symmetry breaking on the surface state of TIs. Finally, we discuss the implications of our findings on applications of magnetically doped topological insulators to spintronics and quantum information technology.

## 2. Methods

The samples investigated in this work consisted of MBE-grown bilayer structures as schematically shown in figure 1(a), where $d_1 = 1, 3, 5, 7$-QL for the 10% Cr-doping level, and $d_1 = 1$-QL for 5% Cr-doping. Hereafter we use the nomenclature $(d_1+d_2)$-$x$% to denote our samples. Details of the growth process, structural characterizations and ARPES studies of these bilayer samples have been reported elsewhere [18,28,29]. Bulk electrical transport measurements on these samples revealed the appearance of anomalous Hall resistance at $T \leq T_c^{3D} \sim 30$ K for $x = 10$% and $\sim 20$ K for $x = 5$%, respectively, as exemplified in figure 1(b). Additionally, surface magneto-optic Kerr effect (SMOKE) measurements have been performed on similar bilayer samples, showing finite Kerr angles $\theta_K$ at $T < T_c^{3D}$, with temperature and magnetic field dependence similar to those of the anomalous Hall resistance [18], as exemplified in supplementary figure 1(a)-(c).

For the STM studies, each bilayer sample was capped with ~ 1 nm Se inside the MBE growth chamber for passivation immediately after the bilayer growth. The sample was subsequently transferred from the growth chamber via a vacuum suitcase to another vacuum chamber, where the sample was annealed at 423 K ~ 473 K for 90 minutes under vacuum (< $10^{-5}$ Torr) to remove the Se capping layer. The exposed bilayer sample was cooled to 300 K in vacuum, and then the sample-containing chamber was filled with Ar gas and loaded into an Ar-filled glove box, where the sample was removed from the chamber and transferred to the STM probe placed in the same glove box. The STM probe was sealed, transferred to its cryostat, and then evacuated down to ~ $10^{-10}$ Torr at liquid helium temperatures. The variable temperature range achievable for our homemade STM system was from 300 K to ~ 10 K, and a superconducting magnet was available to provide magnetic fields up to ~ 7 Tesla.

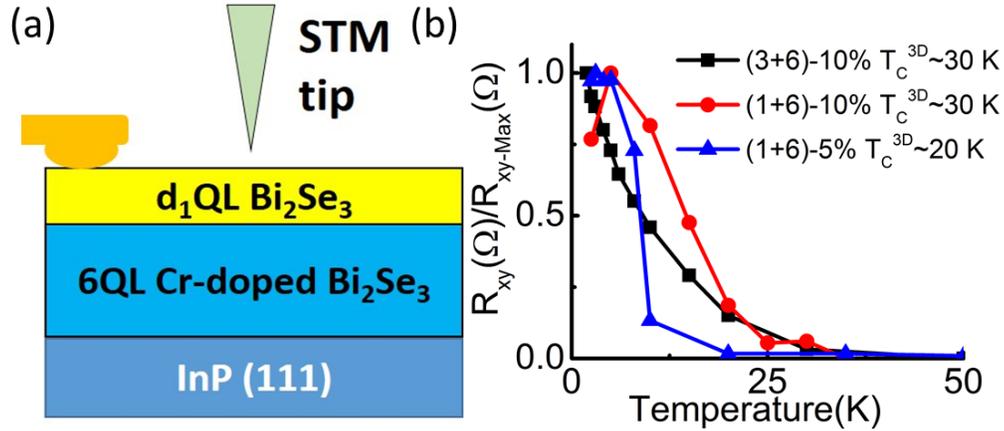

**Figure 1. (a)** Schematics of the side view of a $Bi_2Se_3$ bilayer sample, showing an undoped $Bi_2Se_3$ layer of a thickness $d_1$-QL on top of a Cr-doped $Bi_2Se_3$ layer of a thickness 6-QL grown on InP (111). A gold contact was placed on top of $d_1$. **(b)** Temperature dependent normalized Hall resistance ($R_{xy}$) of the $Bi_2Se_3$ bilayer samples at $H = 0$, with each set of data normalized to its maximum Hall resistance value ($R_{xy\text{-Max}}$). The $R_{xy\text{-Max}}$ values are 1.35 Ω, 1.26 Ω and 0.32Ω for the (3+6)-10%, (1+6)-10% and (1+6)-5% samples, respectively. The $T_c^{3D}$ values are estimated from the temperatures associated with the 1% values of $R_{xy\text{-Max}}$, which yield ~ 30 K, ~ 30 K and ~ 20 K for the (3+6)-10%, (1+6)-10% and (1+6)-5% samples, respectively.

## 3. Results and analysis

In this work both topographic and spectroscopic studies were made on all samples as a function of $T$ (from 300 K to 15 K) and $H$ (from 0 to 3.5 Tesla) using the variable temperature STM.

*3.1. Structural characteristics from surface topographic studies*

The surface topography on large scales revealed pyramid-like terraces with steps corresponding to single atomic layers, as described previously [28,29]. For an averaged, nominal top layer thickness $d_1$-QL, the local thickness of the top layer could vary up to 1-QL. Atomically resolved topographic images exhibited triangular lattice patterns that were always consistent with that of pure $Bi_2Se_3$, as exemplified in figure 2(a) and figure 2(b) for (1+6)-10% and (5+6)-10% samples, respectively. On the other hand, the Fourier transformation (FT) of the surface topography appeared to be dependent on $d_1$. We found that FT of the (1+6)-10% topography showed an expected hexagonal Bragg diffraction pattern for $Bi_2Se_3$ plus an additional, faint superlattice structure (figure 2(c)), which may be attributed to the underlying Cr-doped $Bi_2Se_3$. For instance, a periodic substitution of Cr for Bi as exemplified in figure 2(f) for a two-dimensional projection of the two Bi-layers within one-QL yields a FT pattern (figure 2(e)) similar to that in figure 2(c). This superlattice structure corresponds to a local Cr concentration of 1/12. Another similar structure with a local Cr concentration of 1/8 (figure 2(h)) is also feasible within experimental uncertainties of the superlattice constant and its angle relative to the Bi lattice (figure 2(g)). In contrast, the FT in figure 2(d) for the surface topography of a (5+6)-10% sample only revealed the hexagonal diffraction pattern of pure $Bi_2Se_3$ due to the relatively thick $d_1$ layer. Interestingly, we note that the FT topography of the (1+6)-5% samples also agreed with figure 2(d), suggesting random Cr substitutions of Bi for a smaller Cr concentration, which is consistent with the randomly distributed Cr clusters found with STM studies directly on 2% Cr-doped $Bi_2Se_3$ [30].

*3.2. Zero-field spectroscopic studies*

For the zero-field studies, tunneling conductance ($dI/dV$) *vs.* biased voltage ($V = E/e$) measurements were carried out on each sample over multiple areas, followed by detailed analysis of the spatially resolved spectral characteristics. While apparent spatial variations were found in all samples, systematic investigations led to several general findings. First, all samples revealed gapless Dirac tunneling spectra at 300 K. Second, with decreasing $T$ there were two distinctly different types of spectral characteristics: For samples with nominal $d_1 = 5$ and 7, the tunneling spectra remained gapless for all $T$, except for occasional areas where the actual $d_1$ values in the nominal $d_1 = 5$ sample were ~ 4. In contrast, for samples with nominal $d_1 = 1$ and 3, the majority spectra revealed gapped features at $T < T_c^{2D}$, and the temperature evolution for all samples are exemplified in figure 3 (a)-(d). Third, the surface gap $\Delta(\mathbf{r},T)$, obtained by the spectral analysis illustrated in figure 3(e), appeared to be spatially inhomogeneous where $\mathbf{r}$ denotes the two-dimensional spatial coordinate.

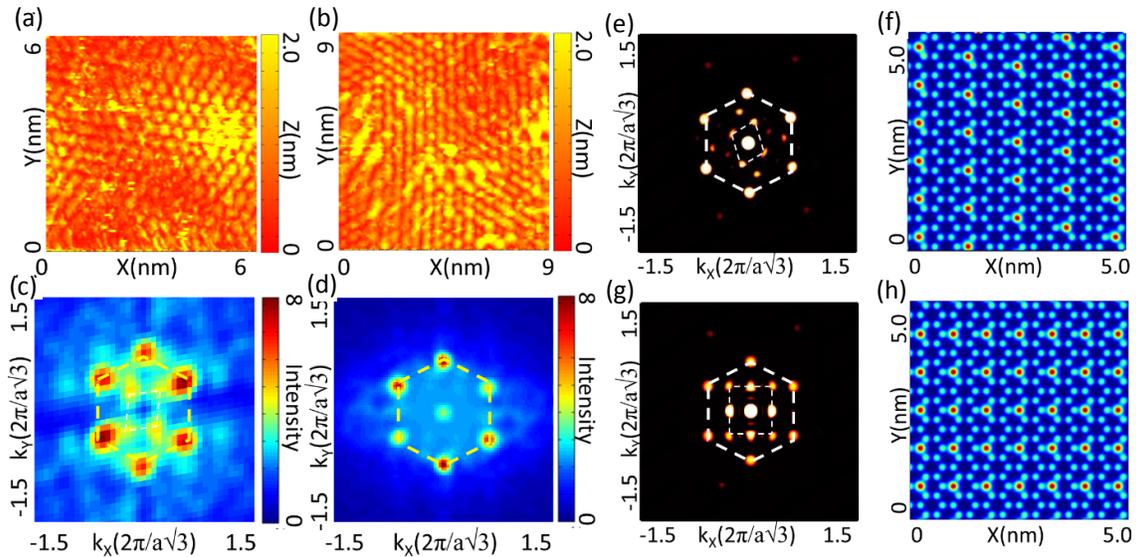

**Figure 2.** Structural characteristics of MBE-grown $Bi_2Se_3$ bilayer samples on InP (111): (a) Surface topography of a (1+6)-10% sample over an area of (6×6) nm$^2$, showing a triangular lattice structure. (b) Surface topography of a (5+6)-10% sample over an area of (6×6) nm$^2$, showing a triangular lattice. (c) Fourier transformation (FT) of the surface topography in a (1+6)-10% sample, revealing a dominant hexagonal reciprocal lattice structure and a secondary superlattice of a much weaker intensity, probably coming from the underlying Cr-doped $Bi_2Se_3$ layer. Here "a" in the reciprocal space scale ($2\pi/a\sqrt{3}$) refers to the in-plane nearest neighbor distance between Bi (Se) and Bi (Se). (d) FT of the surface topography on a (5+6)-10% sample, showing a purely hexagonal reciprocal lattice. (e) Simulated FT of the 1/12 Cr-substituted Bi layer illustrated in (f), showing a FT similar the data in (c). Here the blue dots represent Bi atoms and the red dots represent Cr substitutions. (g) Simulated FT of the 1/8 Cr-substituted Bi layer illustrated in (h), showing a FT also similar the data in (c) within experimental errors.

For a given $\mathbf{r}$, $\Delta$ mostly increased with decreasing $T$ except near $T_x \sim (110\pm10)$ K where a slight dip appeared, and eventually saturated to a maximum value at $T < T_c^{3D} \ll T_c^{2D}$, as exemplified by the

$T$ evolution of the gap maps and the corresponding gap histograms in the left and middle panels of figure 4 (a)-(c) for the (1+6)-5%, (1+6)-10% and (3+6)-10% samples, and also summarized in the right panels for the temperature dependence of the corresponding mean gap $\overline{\Delta}(T)$. Here the mean gap value $\overline{\Delta}$ at a given $T$ was determined from the peak value of Gaussian fitting to the gap histogram, and the errors were determined from the one sigma linewidth of the Gaussian fitting. Based on the data for $\overline{\Delta}(T)$, the onset $T$ for the surface gap opening was found to be $T_c^{2D} = (240 \pm 10)$ K for $x =$ 10% and $T_c^{2D} = (210 \pm 10)$ K for $x = 5\%$, significantly higher than the bulk Curie temperatures $T_c^{3D} =$ 30 ~ 20 K obtained from the onset temperature of the anomalous Hall resistance. This finding of $T_c^{2D} \gg T_c^{3D}$ is in fact consistent with previous reports on other families of 3D-TIs [15-17]. As we shall elaborate further in Discussion, the non-monotonic temperature dependence of the surface gap and the relation of $T_c^{2D} \gg T_c^{3D}$ may be the result of multiple magnetic interaction components in the bilayer samples, with each component having a different temperature dependence and interaction range.

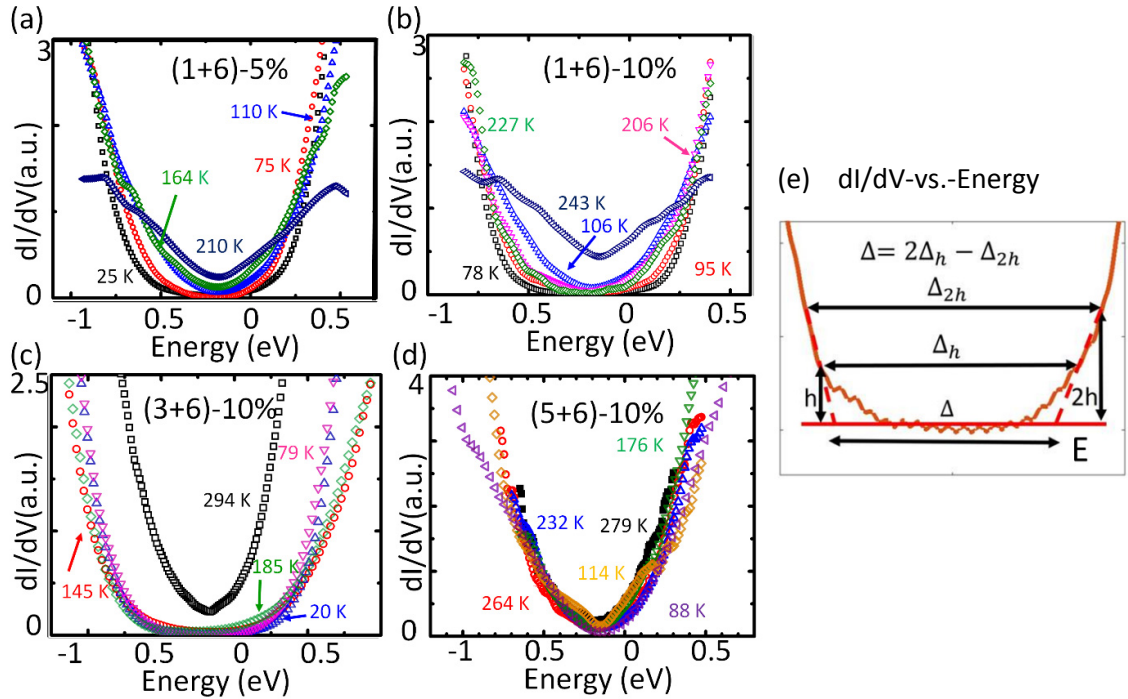

**Figure 3.** Temperature evolution of representative normalized tunneling conductance spectra of $(d_1 + d_2)$-$x\%$ bilayer samples taken at $H = 0$: (a) (1+6)-5%, (b) (1+6)-10%; (c) (3+6)-10%, and (d) (5+6)-10%. For samples with $d_1 = 1$ and 3, each representative spectrum in (a)-(c) at a given $T$ was determined by first taking the spatially resolved tunneling spectra over a fixed area of the sample, plotting the histogram of the gap values to determine the mean gap $\overline{\Delta}$ over this area, and then averaging those spectra with gap values within one sigma of $\overline{\Delta}$. To compensate for the thermal drift of the STM tip and to ensure that the spectral analysis was carried out over the same sample area for all different temperatures, we compared the topographic images taken at all temperatures to identify the overlapped areas. With this procedure, finite gaps were consistently found to develop at low temperatures for samples with $d_1 = 1$ and 3 as exemplified in (a)-(c), whereas all spectra were gapless for samples with $d_1 = 5$ and 7, as exemplified in (d). (e) Schematic illustration showing how the gap is estimated from a realistic normalized tunneling conductance spectrum: Defining the conductance of the inflection point in the tunneling spectrum as $h$ and the corresponding energy difference between the spectral inception points as $\Delta_h$, we identify $\Delta_{2h}$ for the tunneling conductance at $2h$, and then extrapolate an effective gap $\Delta$ at zero conductance from the formula $\Delta = 2\Delta_h - \Delta_{2h}$. The maximum gap thus obtained is consistent with the theoretical values (0.3 ~ 0.5 eV) from the densities of states of $Se^{I,II}$ in Cr-doped $Bi_2Se_3$ [30].

The inhomogeneous gap distribution may be attributed to multiple reasons. First, the Cr-substitution of Bi may not be uniform due to the size mismatch, which could induce lattice strain and inhomogeneous ferromagnetism. Second, the magnetic moments of Cr ions may not be well aligned along the sample c-axis without an external magnetic field. Given that only *c*-axis magnetization component can induce a surface gap in 3D-TIs according to Eqs. (1)-(2), varying spin alignments in different magnetic domains would result in varying surface gaps [31]. Third, the sample surface exhibited terrace structures with ~ 1-QL thickness variations [28], which could give rise to varying proximity-induced gaps from the RKKY interaction mediated by surface Dirac fermions [18,22-24].

To better understand how the aforementioned components contribute to the gap inhomogeneity, we compare the surface topography of the bilayer samples with their corresponding gap maps. As exemplified in supplementary figure 2(a)-(b) and discussed further in the supplementary information, we find that the correlation between a typical surface topography map and the corresponding gap map over the same area of a bilayer sample is finite although relatively weak. The weak correlation is primarily due to small height variations over the typical areas of our STM studies. Thus, the primary cause of spatially inhomogeneous gaps in the bilayer systems may be attributed to inhomogeneous Cr distributions and the misalignment of Cr magnetic moments. Experimentally, the latter effect may be minimized by applying a strong external magnetic field along the c-axis, which is the subject of our investigation in the next subsection.

*3.3. Finite-field spectroscopic studies*

We further investigated the effect of increasing *c*-axis magnetic field ($H$) on the gap distribution over the same area of each sample at a constant *T*. As exemplified in figure 5(a)-(f) for samples of (1+6)-5% and (3+6)-10% taken at $T = 18$ K and $H = 0$, 1.5 T and 3.5 T, the gap maps became increasingly homogeneous and the mean gap value $\bar{\Delta}$ derived from the histogram also increased slightly with increasing $H$. This finding suggests that the observed surface gap is consistent with *c*-axis ferromagnetism induced by Cr-doping and proximity effect, and the characteristic field for saturating the surface state ferromagnetism appears to be larger than that for saturating the bulk ferromagnetism [18], probably due to the helical spin textures of the former. The small but finite residual gap inhomogeneity in high fields may be primarily attributed to spatially inhomogeneous Cr-distributions.

*3.4. Minority spectra*

While the majority of the tunneling spectra in the (1+6)-5%, (1+6)-10% and (3+6)-10% samples revealed gapped characteristics for $T < T_c^{2D}$, spatially localized and intense conductance peaks were occasionally observed along the borders of gapless and gapped regions, as exemplified in figure 6 (a)-(d) for a (1+6)-5% sample and in figure 6 (f) for a (1+6)-10% sample. These long-lived minority spectra either consisted of a single sharp conductance peak at a small negative energy $E = E_-$ near the Dirac point $E_D$; or comprised of double conductance peaks (figure 6(b)) at $E = E_-$ and $E = E_+$, where $E_+$ is near the Fermi energy $E_F = 0$. These double-peak spectral characteristics were consistent with theoretical predictions for magnetic impurity resonances [25]. Further, the numbers of both single- and double-peak impurity resonances at $H = 0$ were found to increase rapidly near $T_c^{2D}$ (figure 6(f)). In contrast, all resonances disappeared under a large c-axis magnetic field at low *T* when gapless regions diminished.

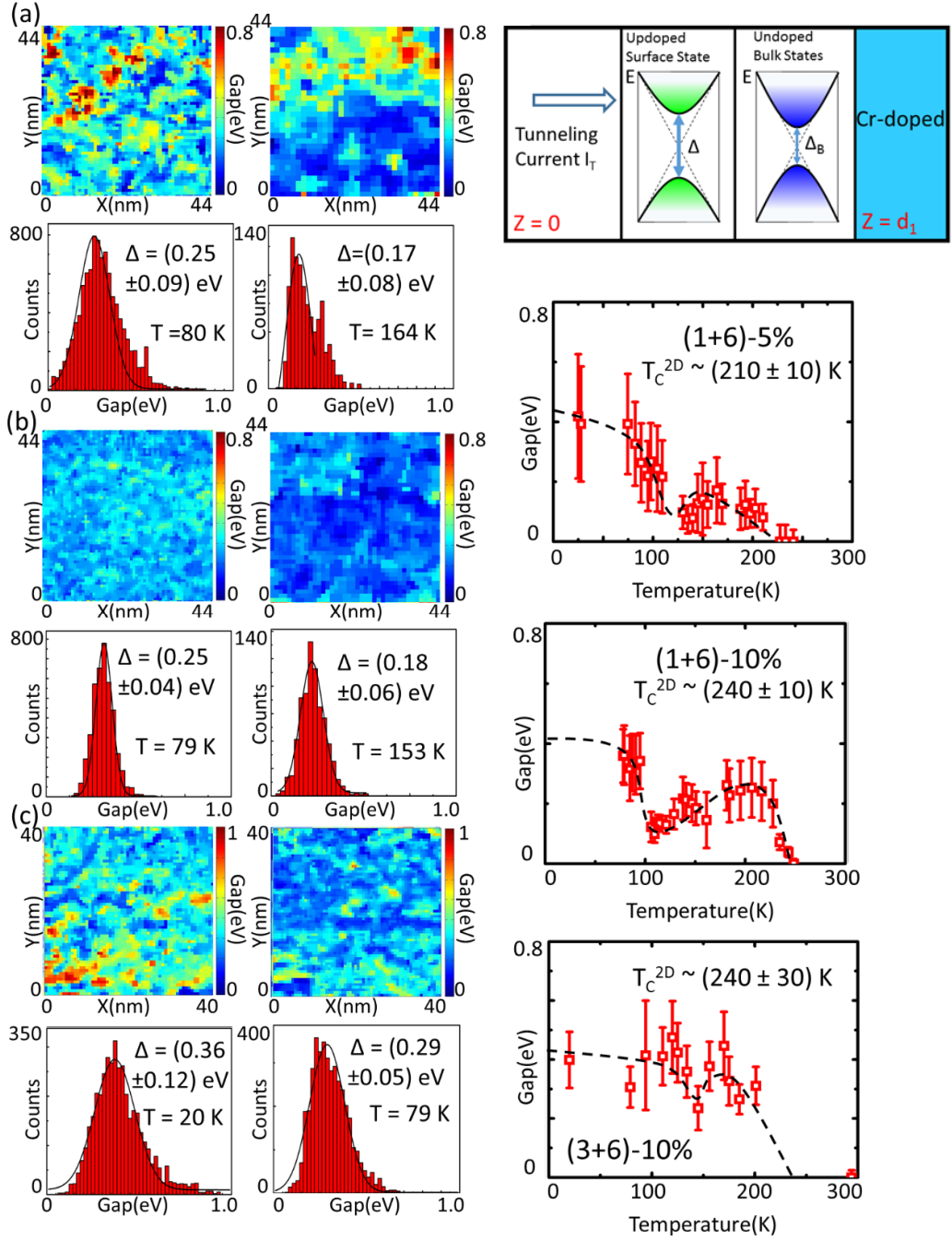

**Figure 4.** Temperature evolution and spatial distribution of the surface gap at $H = 0$: (a) Gap maps and the corresponding histograms of a (1+6)-5% sample taken at $T = 80$ K (left panels) and $T = 164$ K (middle panels), and the $T$-dependence of the mean gap $\bar{\Delta}$ (second right panel). (b) Gap map and the corresponding histograms of a (1+6)-10% sample taken at $T = 79$ K (left panels) and $T = 153$ K (middle panels), and the $T$-dependence of the mean gap $\bar{\Delta}$ (third right panel). (c) Gap maps and the corresponding histograms of a (3+6)-10% sample taken at $T = 20$ K (left panels) and $T = 79$ K (middle panels), and the $T$-dependence of the mean gap $\bar{\Delta}$ (fourth right panel). The first right panel is a schematic illustration of the spatially varying electronic structure experienced by the tunneling current, showing the dominance of the surface state gap $\Delta$ over the bulk gap $\Delta_B$ in determining the measured spectral gap in STS.

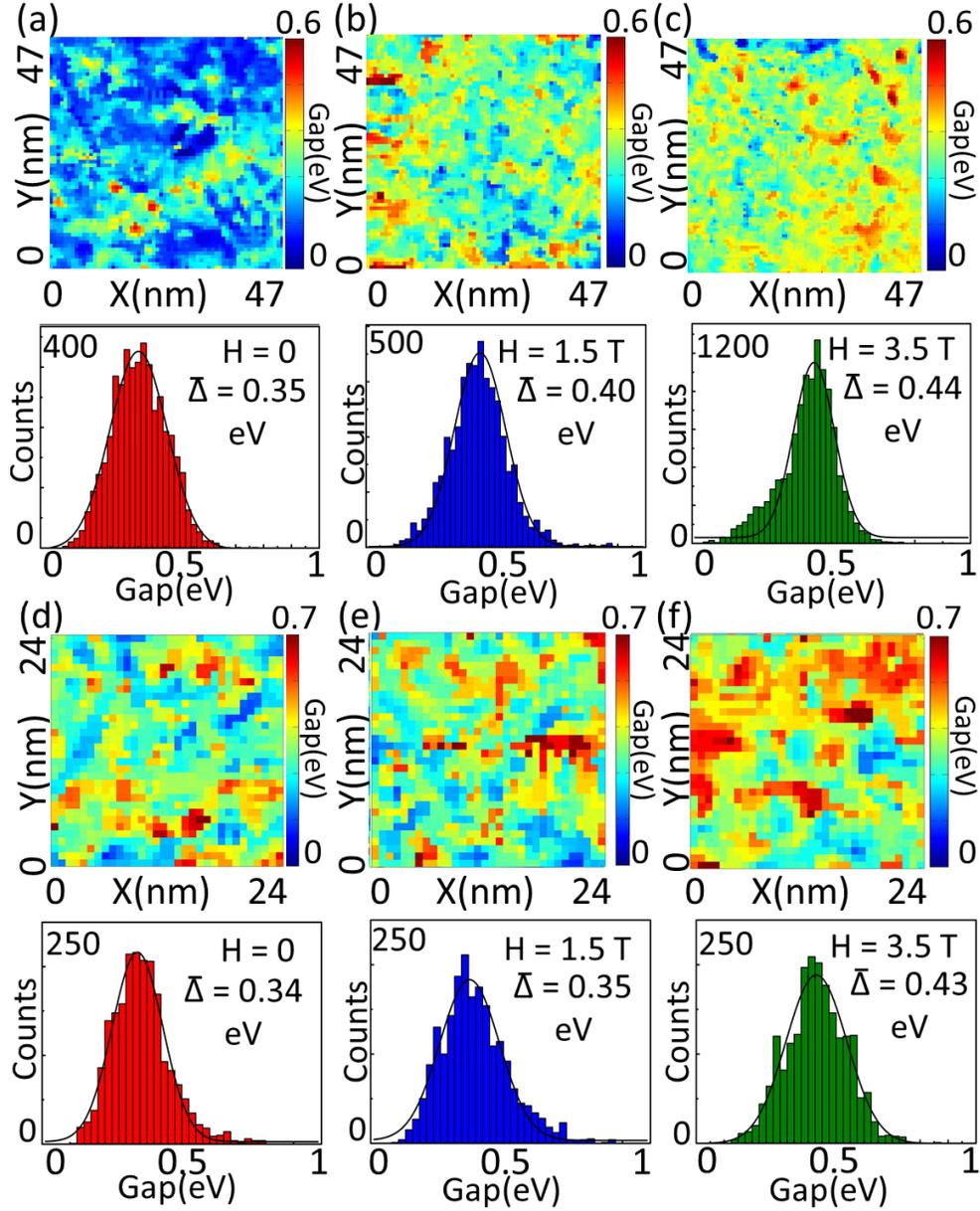

**Figure 5.** Evolution of the surface gap distribution at $T = 18$ K with applied c-axis magnetic field: (a-c) Gap maps (upper panels) and the corresponding gap histograms (lower panels) of a (1+6)-5% sample taken at $H = 0$, 1.5 T and 3.5 T over the same (47 × 47) nm$^2$ area. (d-f) Gap maps (upper panels) and the corresponding gap histograms (lower panels) of a (3+6)-10% sample taken at $H = 0$, 1.5 T and 3.5 T over the same (24 × 24) nm$^2$ area.

We attribute the sharp impurity resonances to isolated Cr-impurities that were far away from neighboring Cr ions and probably had partially diffused from the $d_2$-layer into the interfacial region, because they were only found in zero-field along the borders between gapless and gapped regions, and then disappeared under a large c-axis magnetic field when gapless regions diminished and long-range ferromagnetism was established. The strongly non-monotonic temperature dependence of the number of impurity resonances at $H = 0$ (see figure 6 (f)) may be understood as the result of weakening surface ferromagnetism near $T_c^{2D}$, so that more Cr-impurities became decoupled and acted like isolated impurities. The strong spatial localization and long lifetime of these magnetic impurity resonances at $H = 0$ may be attributed to topological protection of the surface state in 3D-TIs when the Dirac energy $E_D$ is relatively close to $E_F$, similar to the case for non-magnetic impurities [25, 32].

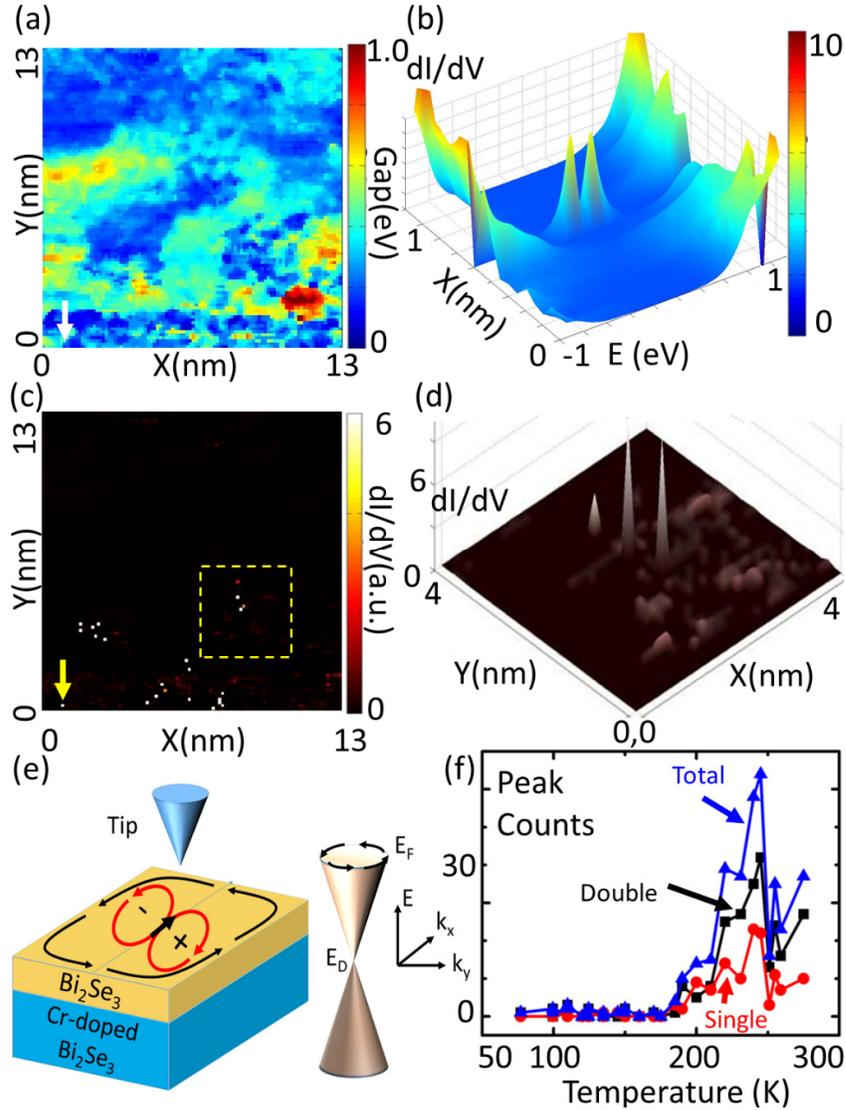

**Figure 6.** Spectral characteristics of isolated magnetic impurities: (a) Gap map of a (1+6)-5% bilayer sample taken at $T = 80$ K and $H = 0$ over an area of $(13\times13)$ nm$^2$, showing both spatially inhomogeneous gaps and some gapless regions (dark blue). The arrow indicates a site where a spatially localized double-resonant spectrum in (b) is observed. (b) Sharp resonances of an isolated impurity are manifested in the $(dI/dV)$ vs. $E$ and $X$ plot, where $X$ is a horizontal linecut across an isolated impurity indicated by the arrow in (c). The sharp double resonant peaks appear at $E = E_-$ and $E = E_+$, where $E_-$ is near $E_D$ and $E_+$ is near $E_F$. (c) Tunneling conductance map taken at $E = E_-$ over the same area as in (a), showing spatially isolated conductance peaks in bright sharp spots. (d) Two-dimensional distribution of the tunneling conductance at bias voltage $V = (E_-/e)$ over a $(4\times4)$ nm$^2$ area indicated by the dashed box in (c), showing three sites with intense impurity resonances. (e) Schematics of a topological defect (red arrows of opposite helicity) due to an isolated magnetic impurity in the surface-state Dirac spin textures (counterclockwise black arrows). (f) $T$-evolution of the counts of single- and double-peak and total impurity resonances over a $(20\times20)$ nm$^2$ area of a $(1 + 6)$-10% sample, showing a rapid increase in the number of impurity resonances near $T_c^{2D}$ for both the single- and double-peak resonances.

We further remark that our finding of atomically sharp impurity resonances due to isolated Cr ions differs from direct STM studies of bare 2% Cr-doped $Bi_2Se_3$ [30]: In the latter case, direct atomic imaging revealed that Cr substituted Bi in clusters so that the resulting spectroscopy did not behave like isolated magnetic impurities embedded in the surface state of pure $Bi_2Se_3$. In contrast, our identification of isolated Cr-impurities in the bilayer systems was solely based on spatially resolved *spectroscopic* evidences without directly imaging the Cr impurities, because Cr ions could not diffuse all the way to the top surface and were still buried by the undoped $Bi_2Se_3$ layer. Therefore, STM *imaging* of all bilayer samples always revealed the atomic structure of pure $Bi_2Se_3$, as exemplified in figure 2 (a) - (b). Additionally, we have found no evidences for the formation of Cr clusters at the interface of our bilayer samples: The STEM studies [18] of typical bilayer samples revealed atomically sharp interfaces without any cluster-induced variations in the c-axis lattice spacing; our bilayer samples also exhibited well defined 3D ferromagnetism below an appreciable $T_c^{3D}$ (20 ~ 30 K), which was consistent with more dispersed and uniform distributions of Cr in order to achieve long-range ferromagnetic order. The latter situation was in stark contrast to the absence of long-range ferromagnetism down to 1.5 K in the 2% Cr-doped $Bi_2Se_3$ sample that exhibited randomly distributed surface Cr-clusters [30]. In fact, a recent report based on direct STM studies on a similar TI system $(Bi_{0.1}Sb_{0.9})_{1.92}Cr_{0.08}Te_3$ (with $T_c^{3D}$ ~ 18 K and 4% Cr-doping) revealed completely dispersed distributions of Cr and inhomogeneous gaps without evidences for cluster formation [33]. These findings are consistent with what we have inferred from our experimental studies of the $(Bi_{1-x}Cr_x)_2Se_3/Bi_2Se_3$ bilayer systems.

## 4. Discussion

In this session we address issues associated with the physical mechanism for proximity magnetism, the physical origin and the magnitude of the surface energy gap, the feasibility of high surface $T_c^{2D}$ values relative to the bulk $T_c^{3D}$, and the topological spin textures in the presence of isolated magnetic impurities.

*4.1. Proximity magnetism in the bilayer system*

Our experimental results have demonstrated that the dependence of the spectral characteristics on $T$, $H$, $x$ and $d_1$ for the bilayer samples are all consistent with the scenario that a surface-state gap was induced by the proximity effect of predominantly *c*-axis ferromagnetism in the Cr-doped bottom layer. The appearance (absence) of gapped tunneling spectra below $T_c^{2D}$ for bilayer samples with $d_1 \leq$ 3-QL ($d_1 \geq$ 5-QL) suggests that the proximity effect due to *c*-axis magnetic correlation is limited to a critical thickness $d_c$ ~ 4-QL.

In our investigation of the Cr-doped bilayer system, it is important to realize that the undoped $Bi_2Se_3$ layer should *not* be considered as an isolated ultrathin film on a "substrate" of Cr-doped $Bi_2Se_3$. This realization is because of the continuous and seamless growth of undoped $Bi_2Se_3$ layer on top of the isostructure, Cr-doped $Bi_2Se_3$ layer [18], leading to an effective total thickness of ($d_1+d_2$)-QL. In contrast, for ultra-thin TI films on dissimilar substrates, an energy gap can open up due to coupling between the top and bottom topological surface states, and a Rashba-like coupling with further energy splitting in momentum can also be expected due to the asymmetric chemical potentials between the surface of the thin film and its interface with a dissimilar substrate [34]. This Rashba-like coupling has indeed been confirmed by ARPES and STS studies for 2-QL to 5-QL $Bi_2Se_3$ on various dissimilar substrates [35, 36]. In particular, the energy gap induced by the Rashba-like coupling for ultra-thin TI films on dissimilar substrates does *not* exhibit any discernible temperature or magnetic field dependence, which is in sharp contrast to the behavior of the proximity-magnetism induced surface gaps described in this work.

Having established proximity magnetism as the physical origin of the surface gap in the bilayer system, we consider next the primary components that contribute to the surface ferromagnetism in the undoped $Bi_2Se_3$ layer of $d_1 \leq$ 3-QL. A natural component is the dipole fields associated with the magnetic moments of Cr ions. Additionally, RKKY-like magnetic interactions mediated by the bulk carriers and the surface Dirac fermions are also feasible contributions to the surface ferromagnetism. Therefore, we may express the effective ferromagnetic coupling constant $J_{eff}$ in terms of the sum of these three components, $J_{eff} = (J_{dipole} + J_{bulk}) + J_{Dirac}$, where $(J_{dipole} + J_{bulk}) \equiv J_{3D}$ may be considered as a 3D component and $J_{Dirac} \equiv J_{2D}$ a 2D component. We expect $J_{3D}$ to vary slowly with $d_1$ because it primarily represents the bulk magnetic properties of the Cr-doped $Bi_2Se_3$. In contrast, $J_{2D} \propto \exp(-2d_1/d_c)$ is strongly dependent on $d_1$ because it involves the wavefunction overlaps between the surface Dirac fermions of the undoped $Bi_2Se_3$ layer and those of the $d$-electrons in Cr-doped $Bi_2Se_3$ layer [18, 22-24].

Concerning the temperature dependence of $J_{eff}$, we expect $J_{3D}$ to dominate at low $T$ because of the strong confinement of Dirac fermions to the surface layer and therefore negligible RKKY interaction between surface Dirac fermions and bulk $d$-electrons. On the other hand, $J_{3D}$ is sensitive to 3D long-range order of magnetic moments and so should diminish significantly above $T_c^{3D}$. In contrast, finite temperature can enhance wavefunction overlaps between the surface Dirac fermions and the interfacial $d$-electrons. Therefore, we expect $J_{2D}$ to dominate at elevated temperatures, and have further elaborated our rationale for this temperature dependence in Supplementary Information. Moreover, the long Fermi wavelength of surface Dirac fermions could result in much enhanced RKKY interaction in the bilayer system so that $J_{2D} > J_{3D}$ [18, 22-24], which may account for the empirical observation of $T_c^{2D} >> T_c^{3D}$.

*4.2. The magnitude and temperature dependence of the surface gap*

Based on the scenario described above and noting that the proximity-induced surface gap is given by $\Delta = (J_{eff}M)$ where $M$ is the surface magnetization that increases monotonically below $T_c^{2D}$, we find that with decreasing $T$, the surface gap $\Delta$ ($T < T_c^{2D}$) can indeed exhibit a non-monotonic $T$-dependence that is consistent with our experimental findings depicted in figure 4. As qualitatively exemplified in supplementary figure 3, $\Delta$ ($T$) first increases with decreasing $T$ for $T < T_c^{2D}$, and then exhibits a "dip" at $T_x$ where $T_c^{2D} > T_x > T_c^{3D}$, and $T_x$ represents a dimensional crossover temperature below which $J_{2D}$ becomes negligible. While we do not have sufficient information to model $\Delta(T)$ quantitatively, the qualitative agreement of supplementary figure 3 with experimental results in figure 4 suggests that our proposed scenario for proximity magnetism in the bilayer system may provide useful guide for future investigation.

Next, we comment on the magnitude of the proximity-induced surface gaps in $Bi_2Se_3$, which were found to be comparable or even larger than the bulk gap ($\sim$ 0.3 eV) of $Bi_2Se_3$. While one may question whether it is feasible for a magnetism-induced surface gap to exceed its bulk gap of a topological insulator, we point out that the bulk gap $\sim$ 0.3 eV of $Bi_2Se_3$ is an indirect gap; the direct gap at the $\Gamma$-point where the Dirac cone resides is in fact $>$ 0.5 eV based on bandstructure calculations [37]. In this context, the largest mean gap values (0.4 eV $\sim$ 0.5 eV) found in our work are well within the range of the bulk gap. We further note that the gap values determined from STS studies would be dominated by the local gap of the surface layer, whereas the gap value determined from ARPES would be sensitive to the smallest gap over an extended surface area. Therefore, the surface gap determined from STS tends to be larger than that determined from ARPES for an inhomogeneous sample. Indeed, ARPES studies on $Bi_{1.8}Cr_{0.2}Se_3$ have revealed a surface gap of $\sim$ 0.2 eV [18], which is slightly smaller than but still in reasonable agreement with our finding of $\Delta = (0.25 \pm 0.09)$ eV for a (1+6)-10% sample at $H = 0$ and $T = 79$ K, as shown in Fig. 4(b).

*4.3. The feasibility of a much higher surface $T_c^{2D}$ than the bulk $T_c^{3D}$*

Concerning the physical feasibility of a relatively high $T_c^{2D}$, we note that the energy splitting of the double-peak spectrum associated with isolated magnetic impurities in the surface state is comparable to the effective ferromagnetic coupling constant $J_{eff}$ [21]. Noting that the mean-field Curie temperature may be estimated from $J_{eff}$ via the relation $k_B T_c^{2D} \sim S(S+1)(J_{eff}/3)$ [38] and that $J_{eff} \sim 0.1$ eV as determined from the measured energy splitting of the double resonances, we obtain $T_c^{2D} \sim 250$ K by assuming $S = 1/2$. While this rough estimate does not include the screening effect from Dirac fermions, it is still comparable to our STS measurements of $T_c^{2D} \sim 240$ K for $x = 10\%$ and $T_c^{2D} \sim 210$ K for $x = 5\%$ for the observed surface ferromagnetism, and is much higher than the bulk $T_c^{3D} \sim 25$ K. The disparity of $T_c^{2D}$ and $T_c^{3D}$ may be attributed to the different microscopic mechanisms for mediating ferromagnetism via the surface-state Dirac fermions versus via the bulk-state carriers, particularly given the diverging Fermi wavelength of the Dirac fermions when the Fermi energy approaches the Dirac point [18, 22-24]. While quantitative details of the microscopic mechanism responsible for $T_c^{2D} \gg T_c^{3D}$ require further investigation, the relatively high $T_c^{2D}$ values are promising for realistic spintronic applications, particularly if our findings may be generalized to more homogeneously Cr-doped 3D-TIs such as $(Bi_{1-x}Cr_x)_2Te_3$ [18].

*4.4. Topological spin textures*

The appearance of both single- and double-peak magnetic impurity resonances is suggestive of two types of topological defects associated with an isolated magnetic impurity: The single-peak resonance may be attributed to a helical Dirac spin texture coupling to a magnetic moment pointing along the *c*-axis, whereas the double-peak resonance is associated with two opposite chiral spin textures of Dirac fermions coupling to an in-plane magnetic moment [21], as illustrated in figure 6 (e). Our assignment of the single-peak resonances in this work to the helical Dirac spin texture coupled with isolated *c*-axis magnetic moments can be justified by their strong and non-monotonic temperature dependence near $T_c^{2D}$ (figure 6 (f)), and also by their suppression under large *c*-axis magnetic fields. In contrast, our previous studies of non-magnetic impurity resonances in pure $Bi_2Se_3$ MBE-grown thin films [32] did not find any dependence of their occurrences on either the temperature or the applied magnetic field, implying that the physical origin of the single-peak resonances in pure $Bi_2Se_3$ is fundamentally different from that of the single-peak resonances observed in the bilayer systems.

Our finding of occurrences of impurity resonances only along the borders of gapped and gapless regions is consistent with suppressed long-range magnetic order along the boundaries between the *c*-axis oriented (gapped) and in-plane oriented (gapless) magnetic domains [31]. Thus, proximity-induced magnetization in the top $Bi_2Se_3$ layer is also much suppressed for regions above the boundaries of *c*-axis and in-plane magnetic domains [31], and Cr-ions located in these regions were more likely to decouple from long-range magnetization and became effectively isolated magnetic impurities, particularly with the assistance of thermal fluctuations at sufficiently high temperatures (figure 6 (f)).

Finally, we note that the spin textures associated with an in-plane magnetic impurity may be considered as a stable "topological bit" with two levels associated with the two opposite spin chirality. In principle, for a completely isolated topological bit, the two-level states may be tuned by a local *c*-axis magnetic field. Further, for $E_F \rightarrow E_D$ these topological bits are long-lived as the result of topological protection. On the other hand, tuning $E_F$ away from $E_D$ will result in increasing coupling among spatially separated topological bits [21], which may be viewed as inducing an effective entanglement of wavefunctions among these topological bits. The feasibility of tuning the two levels within a topological bit by an external magnetic field and the coupling among spatially separated bits by gating the Fermi level may present an interesting opportunity for applying these topological bits to quantum information technology.

## 5. Conclusion

We have demonstrated scanning tunneling spectroscopic evidences for magnetism-induced massive Dirac spectra and topological defects in the surface state of MBE-grown 3D-TIs that consisted of bilayers of undoped $Bi_2Se_3$ on top of Cr-doped $Bi_2Se_3$. For an undoped layer thinner than or equal to 3-QL, proximity-induced ferromagnetism was manifested by the appearance of a surface gap Δ below a transition temperature $T_c^{2D} > 200$ K, and the value of $T_c^{2D}$ was found to be Cr-doping ($x$) dependent and also much higher than the bulk ferromagnetic transition temperature $T_c^{3D}$ (<~ 30 K). Our finding of $T_c^{2D} >> T_c^{3D}$ may be attributed to the much longer range of the RKKY interaction mediated by the surface Dirac fermions relative to that mediated by the bulk itinerant carriers. The gap distribution was spatially inhomogeneous, and its average value and spatial homogeneity at low $T$ increased with $H$ and $x$, confirming its magnetic origin. On the other hand, the non-monotonic $T$-dependence of Δ may be attributed to proximity magnetism induced by two types of contributions: a 3D contribution from the bulk magnetism that dominates at low $T$, and a 2D contribution associated with the RKKY interactions mediated by surface Dirac fermions, which dominates at $T_c^{3D} << T < T_c^{2D}$. The appearance of a gap below $T_c^{2D}$ is the result of global time-reversal symmetry breaking in the surface state of topological insulators. This global effect is in contrast to the local time reversal symmetry breaking associated with isolated magnetic impurities in the surface state, which tends to induce excess densities of states near the Dirac point. Finally, topological defects associated with helical spin textures of Dirac fermions around isolated magnetic impurities were manifested as long-lived sharp double-peak spectral resonances along the borders of gapped and gapless regions. The strong spatial localization and long life-time of these magnetic impurity resonances represent interesting topological bits that are potentially useful for applications to quantum information technology.

## Acknowledgement

This work at Caltech was jointly supported by DARPA and NSF. The work at UCLA was supported by DARPA. We thank Professors Gabriel Aeppli, Alexei Kitaev, Nai-Phuan Ong and Thomas Rosenbaum for useful discussions.

# Magnetism-induced massive Dirac spectra and topological defects in the surface state of Cr-doped Bi$_2$Se$_3$-bilayer topological insulators

C.-C. Chen[1,2], M. L. Teague[1,2], L. He[3], X. Kou[3], M. Lang[3], W. Fan[1], N. Woodward[1], K.-L. Wang[3] and N.-C. Yeh[1,2,4*]

[1] Department of Physics, California Institute of Technology, Pasadena, CA 91125, USA
[2] Institute of Quantum Matter and Information, California Institute of Technology, Pasadena, CA 91125, USA
[3] Department of Electrical Engineering, University of California, Los Angeles, CA 90095, USA
[4] Kavli Nanoscience Institute, California Institute of Technology, Pasadena, CA 91125, USA
[*] E-mail: ncyeh@caltech.edu

## Supplementary Information

The supplementary information includes the following figures and related information:

Supplementary Figure 1: SMOKE measurement of Cr-doped monolayer and bilayer topological insulators

Supplementary Figure 2: Correlation between the surface topography and the gap maps of the Cr-doped TI bilayer samples

Supplementary Figure 3: Temperature dependent 2D and 3D contributions to the proximity-induced surface gap in Cr-doped TI bilayer systems

1. **Surface Magneto-Optic Kerr Effect (SMOKE) measurements of Cr-doped monolayer and bilayer topological insulators**

Details of the experimental setup and data analysis of the SMOKE measurements have been described in *Nano Lett.* **13**, 4587 (2013). Briefly, the experimental setup used elliptical polarized reflected light to probe the magnetic properties of materials. The SMOKE system was integrated with a temperature-controlled cryostat to achieve cryogenic temperatures down to 2.8 K.

The SMOKE measurements involve spatially averaged signals over large areas (10's ~ 100's of $\mu m^2$), which are about *five to six orders of magnitude larger* than the typical areas investigated by our local-probe STM studies. Because of the large-area averaged effect, finite SMOKE signals can only arise from non-vanishing net ferromagnetic signals over the areas studied. Therefore, for randomly oriented magnetic domains plus inhomogeneity within each magnetic domains as seen by our local-probe STM experiments at $T < T_c^{2D}$, the SMOKE signal would essentially average to zero for $T_c^{3D} < T < T_c^{2D}$. On the other hand, when $T < T_c^{3D}$, clear manifestation of the Kerr angle $\theta_K$ and the anomalous Hall resistance have been verified, and these results have also been discussed in detail in *Nano Lett.* **13**, 4587 (2013). Here we include experimental results taken on a similar bilayer system (with $T_c^{3D} \sim 25$ K and the magnetoresistance and anomalous Hall resistance results all consistent with the $Bi_2Se_3/Bi_{2-2x}Cr_{2x}Se_3$ bilayer system studied in this work) of $d_1$-thickness of $(Bi_{0.5}Sb_{0.5})_2Te_3$ ($d_1 = 0, 3, 6$-QL) on top of 6-QL $Cr_{0.08}(Bi_{0.57}Sb_{0.39})_2Te_3$, as illustrated below in Supplementary Figure 1.

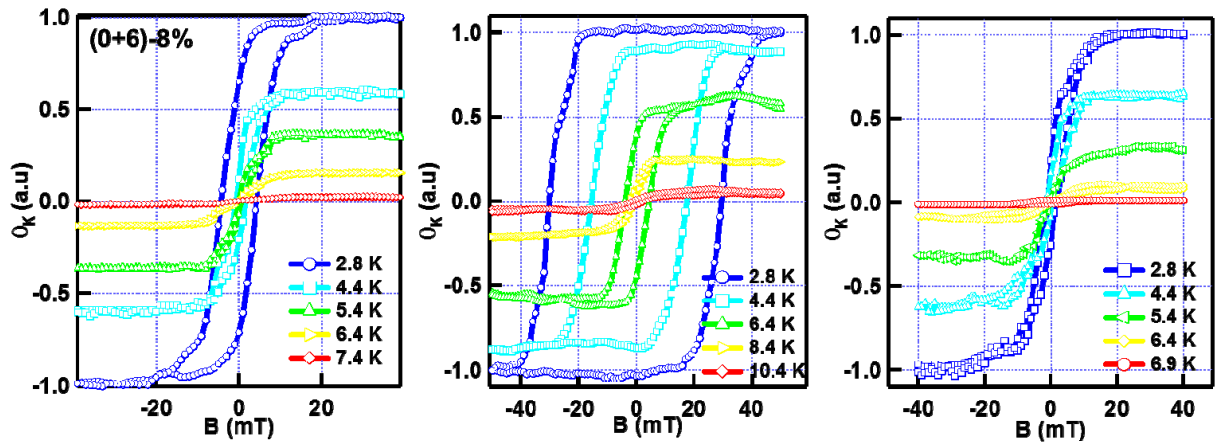

**Supplementary Figure 1| Temperature-dependent SMOKE measurement of Cr-doped monolayer and bilayer topological insulators:** (a-c) Out-of-plane magnetizations (reflected in $H_c$) measured by a polar-mode SMOKE setup. All of the modulation-doped samples have the same 6-QL $Cr_{0.16}(Bi_{0.54}Sb_{0.38})_2Te_3$ bottom $d_2$-layer, with different top $d_1$-layer thicknesses (with $d_1 = 0$, 3 and 6-QL). The largest magnetization is produced in the 3-QL TI/6-QL Cr-doped TI structure, which is consistent with the transport measurements and has been explained by a model based on RKKY interaction mediated by surface Dirac fermions, as discussed in detail in *Nano Lett.* **13**, 4587 (2013).

## 2. Correlation between the surface topography and the gap maps of the Cr-doped TI bilayer samples

The spatially inhomogeneous surface gap found in Cr-doped TI bilayer structures with $d_1 \leq 3$-QL has been attributed to three factors, including inhomogeneous Cr distributions, unaligned magnetic moments of Cr, and height variations in the MBE-grown TI bilayer samples as the result of terrace-like structures.

To investigate how much the height variation may influence the surface gap distribution, we directly compare the topographic map and the gap map of a Cr-doped bilayer sample, (3+6)-10%, as shown in Supplementary Figure 2. We calculate the mean of the correlation function $\langle C \rangle$ between the two maps by the following relation:

$$\langle C \rangle \equiv \frac{\sum_{m,n}(A_{mn}-\bar{A})(B_{mn}-\bar{B})}{\left[\sum_{m,n}(A_{mn}-\bar{A})^2\right]\left[\sum_{m,n}(B_{mn}-\bar{B})^2\right]}.$$

Here the subscripts $m$, $n$ refer to the pixel indices of the "$A$" and "$B$" maps that respectively correspond to the surface topography and the surface gap taken on the same sample and in the same area, whereas $\bar{A}$ ($\bar{B}$) denotes the mean value of the entire surface topography (surface gap) map. We found that the maps shown in Supplementary Figure 2 yield $\langle C \rangle \approx 0.08$, which corresponds to weak correlation. In general, $\langle C \rangle = 1$ implies perfect correlation, $-1$ implies perfect anti-correlation, and 0 implies no correlation.

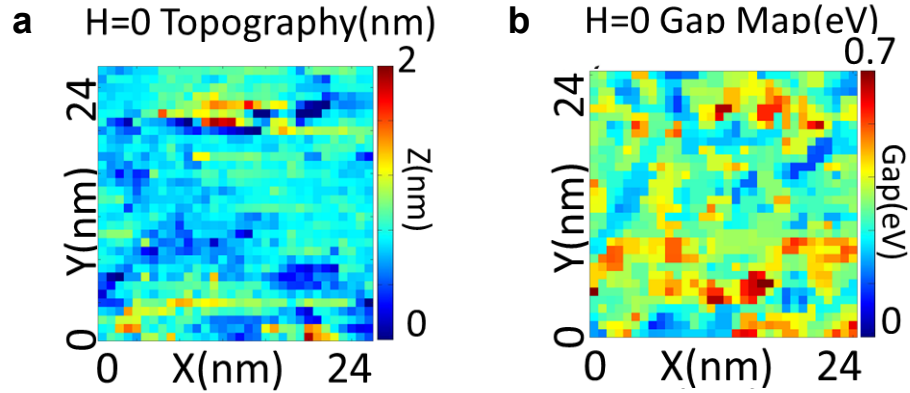

**Supplementary Figure 2| Comparison of the surface topography and the gap map of the same area in a bilayer sample (3+6)-10% at H = 0:** (a) Surface topography over a (24×24) nm² area. (b) The gap map of the same area as in (a). The correlation $\langle C \rangle$ calculated using the relation specified above for the two maps is found to be 0.08, suggesting weak correlation.

# 3. Temperature dependent 2D and 3D contributions to the proximity-induced surface gap in Cr-doped TI bilayer systems

We assume that the proximity-induced surface gap is given by $\Delta = (J_{\text{eff}} M)$ where $M$ is the surface magnetization that increases monotonically below $T_c^{2D}$, and $J_{\text{eff}}$ is the ferromagnetic coupling constant with both contributions from the 3D components $(J_{\text{dipole}} + J_{\text{bulk}}) \equiv J_{3D}$ and the 2D component $J_{\text{Dirac}} \equiv J_{2D}$ associated with surface Dirac fermion-mediated RKKY interaction. As discussed in the manuscript and further elaborated in the end of this section, we expect $J_{3D}$ to dominate at low $T$ and $J_{2D}$ to dominate at elevated temperatures. Moreover, the long Fermi wavelength of surface Dirac fermions could result in much enhanced RKKY interaction in the bilayer system so that $J_{2D} > J_{3D}$.

Based on the aforementioned assumption, the surface gap $\Delta$ $(T < T_c^{2D})$ is expected to exhibit a non-monotonic $T$-dependence, which is consistent with our experimental findings depicted in figure 4: $\Delta (T)$ first increases with decreasing $T$ for $T < T_c^{2D}$, and then exhibits a "dip" at $T_x$ where $T_c^{2D} > T_x > T_c^{3D}$, and $T_x$ is a dimensional crossover temperature below which $J_{\text{Dirac}}$ becomes negligible. While we do not have sufficient information to model $\Delta (T)$ quantitatively, we demonstrate in Supplementary Figure 3 how the qualitative $T$-dependence of $J_{\text{eff}}$ and $M$ proposed in Subsection 4.1 of the manuscript yields non-monotonic $\Delta$-vs.-$T$ behavior similar to the empirical results in figure 4.

In order to qualitatively emulate the aforementioned $T$-dependence of $J_{2D}$, $J_{3D}$, $J_{\text{eff}}$ and $M$, we employ the following functional forms

$$\Delta(T) = J_{\text{eff}}(T) M(T);$$

$$M(T < T_c^{2D}) = M_0 \sqrt{1 - (T/T_c^{2D})}; \quad M(T > T_c^{2D}) = 0;$$

$$J_{\text{eff}}(T) = J_{2D}(T) + J_{3D}(T);$$

$$J_{2D}(T) = J_{2D}^0 \exp\left[-\left(\frac{T_c^{2D}}{T}\right)^{3/2}\right];$$

$$J_{3D}(T) = J_0 + J_{3D}^0 \exp\left[-\left(\frac{T}{T_c^{3D}}\right)^2\right].$$

Using the above functional forms and the parameters:

$$M_0 = 1, \quad J_{2D}^0 = 1, \quad J_{3D}^0 = 0.315, \quad J_0 = 0.1,$$

we find that for $T < T_x$, the $T$-dependence of $J_{\text{eff}}$ is largely determined by that of $J_{3D}$, whereas for $T_x < T < T_c^{2D}$, the $T$-dependence of $J_{\text{eff}}$ is largely determined by that of $J_{2D}$. Further, the surface gap also exhibits a non-monotonic $T$-dependence with a minimum at $T_x$.

Although the functional forms given above can produce a non-monotonic $T$-dependence of $\Delta$ that is in reasonable agreement with empirical results, we caution that one should not over-interpret the physical significance of the specific functional forms given above because of too many parameters that can only be determined from knowing the microscopic details. Further theoretical studies will be necessary to fully understand the temperature dependence of the proximity-induced surface gap.

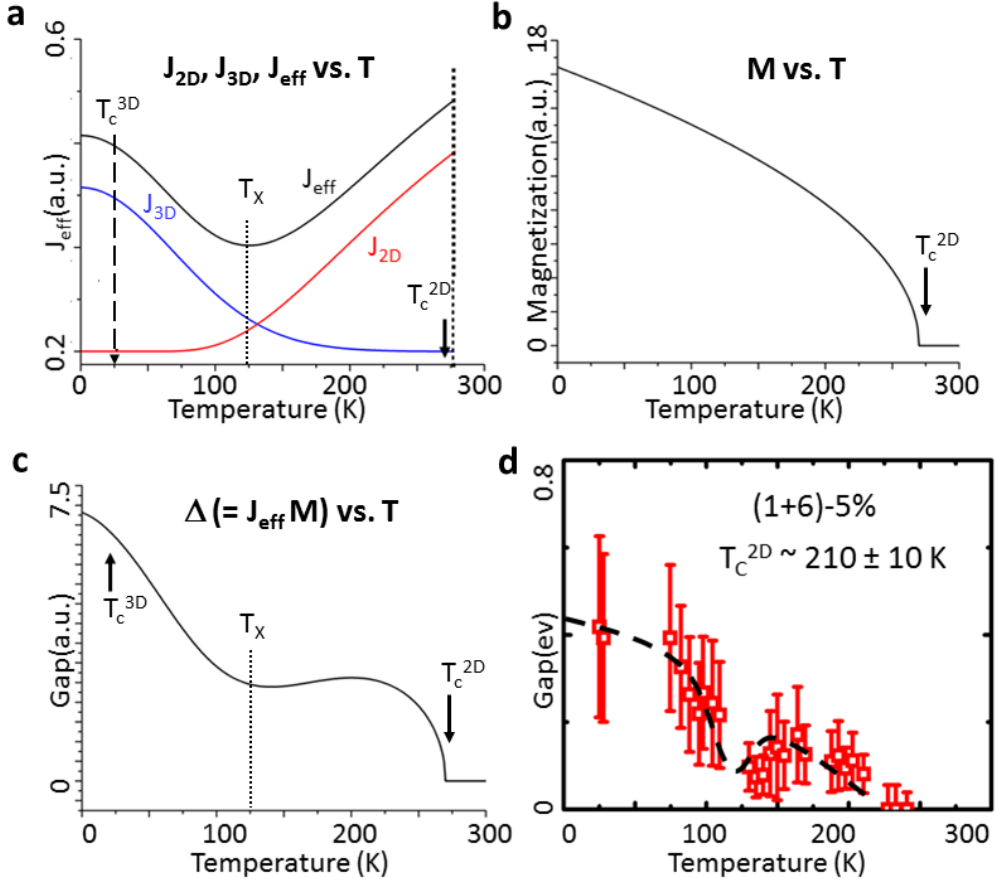

**Supplementary Figure 3| Temperature (T) dependent 2D and 3D contributions to the proximity-induced surface gap in the Cr-doped TI bilayer system.** (a) Schematic temperature ($T$) dependence of the ferromagnetic coupling constant $J_{\text{eff}}$ that consists of a 3D component $J_{3D}$ that dominates at low $T$, and a 2D component $J_{2D}$ that dominates at higher temperatures. The specific functional forms for $J_{3D}$ and $J_{2D}$ are given in the text. (b) Schematic $T$-dependence of the surface magnetization $M$, with the specific functional form specified in the text. (c) The surface gap $\Delta = J_{\text{eff}} M$ reveals a non-monotonic $T$-dependence, consistent with our experimental findings shown in figure 4 and reproduced in (d) for a (1+6)-5% sample.

Finally, we discuss our rationale for a dominant $J_{2D}$ at elevated temperatures. First of all, we note that the wavefunction of Dirac fermions may be expressed as a linear combination of orthonormal eigenfunctions of the Hamiltonian. While the eigenfunctions are independent of temperature, the wavefunction distribution among the eigenfunctions is certainly a function of temperature. In addition, the RKKY interaction associated with $J_{2D}$ must involve spatially extended electronic wavefunctions along the c-axis, and the spatial extension unavoidably involves multiple atomic layers so that thermally assisted processes (*e.g.*, assisted by phonons or impurity states in the bulk) similar to the case of variable-range-hopping conductivity are likely. In this context, higher temperatures could effectively reduce the hopping barrier and enhance the RKKY interaction mediated by the surface Dirac fermions. More specifically, we note that at elevated temperatures the electronic wavefunctions of surface Dirac fermions will involve not only the ground-state eigenfunction (which is confined to 2D) but also the excited-state eigenfunctions (which could extend beyond 2D). Therefore, any interaction matrix element involving the surface Dirac fermions will also be temperature dependent. In particular, if the interaction involves phonon-assisted processes, more significant temperature dependence may be expected.

Empirically, there are two pieces of evidences for strong temperature dependence in the electronic contributions of Dirac fermions, which could be considered as indirectly support for our conjecture of temperature dependence in the 2D RKKY interaction mediated by Dirac fermions. The first piece of evidence was associated with the observation of quantum anomalous Hall effect (QAHE) by C-Z Chang *et al.* in *Science* **340**, 167 (2013): While theoretical predictions for the occurrence of QAHE based on purely topological nature of the magnetic topological insulators would have only required the sample temperature to be sufficiently lower than the Curie temperature ~ 15 K, empirically QAHE could only be observed at $T \sim 30$ mK; slight increase of the temperature to $T \sim 90$ mK already caused significant deviations from the ideal behavior of QAHE. Although there has been no theoretical account for this unexpected temperature dependence of the QAHE in topological insulators, the empirical finding strongly suggests the relevance of temperature to the electronic contributions of Dirac fermions so that increasing temperature could change the 2D characteristics of Dirac fermions in the ground state to more 3D like in the excited states, thus diminishing the QAHE for a strictly 2D system.

The second piece of evidence came from our detailed temperature dependent studies of topological insulator films [L. He *et al.*, Nano Letters **12**, 1486 (2012)], which revealed that the electrical conductivity of topological insulators was dominated by the surface conductance only at sufficiently low temperatures; at intermediate temperatures both surface and bulk contributions were relevant, and then at high temperatures bulk effects dominated. Thus, the electronic contributions of the surface Dirac fermions in topological insulators appear to be strongly dependent on the temperature, which imply that the 2D RKKY interaction mediated by the surface Dirac fermions in our bilayer system would also be temperature dependent.

In summary, to achieve proximity-induced surface ferromagnetism in our bilayer systems through the 2D RKKY interaction mediated by the surface Dirac fermions, the electronic wavefunctions of the surface Dirac fermions must extend beyond the surface layer. While the mechanism(s) that mediate the spread of wavefunctions are not clear yet, possible physical mechanisms such as phonon and/or impurity-states assisted hopping are all strongly temperature dependent. Empirically it has also been shown that the electronic contributions of surface Dirac fermions to such phenomena as the QAHE and electrical conductivity reveal strong temperature dependence, suggesting that the wavefunctions of surface Dirac fermions in the excited states differ significantly from those in the ground state. Consequently, it is feasible to suggest that the magnetic coupling coefficient $J_{2D}$ that involves the surface Dirac fermions is also temperature dependent, with negligible contributions at low temperatures due to strongly surface-confined electronic wavefunctions in the ground state and thus diminished wavefunction overlaps with the underlying Cr atoms.